\documentstyle[12pt]{article}
\newcommand{\be}{\begin{equation}}
\newcommand{\ee}{\end{equation}}
\newcommand{\ba}{\begin{eqnarray}}
\newcommand{\ea}{\end{eqnarray}}

\renewcommand{\thefootnote}{\fnsymbol{footnote}}

\begin{document}
\setlength{\baselineskip}{.7cm}
\renewcommand{\thefootnote}{\fnsymbol{footnote}}
\sloppy

\begin{center}
\centering{\bf \Large General theory of the modified Gutenberg-Richter law\\ for large seismic moments}
\end{center}
\vskip 1cm

\begin{center}
\centering{
Didier Sornette$^{1,2}$ and Anne Sornette$^2$\\
$^1$  Institute of Geophysics and Planetary Physics \\and
Department of Earth and Space Sciences\\
University of California, Los Angeles, California 90095\\
$^{2}$ Laboratoire de Physique de la Mati\`ere Condens\'ee\\
CNRS and Universit\'e de Nice-Sophia Antipolis\\ Parc
Valrose, 06108 Nice Cedex 2, France\\
}
\end{center}

{\bf Abstract} \\
It is well-known that the Gutenberg-Richter power law distribution has to be modified
for large seismic moments, due to energy conservation and geometrical reasons. Several
models have been proposed, either in terms of a second power law 
with a larger $b$-value beyond a cross-over
magnitude, or based on a ``hard'' magnitude cut-off or a ``soft'' 
magnitude cut-off using an exponential taper.
Since the large scale tectonic deformation is dominated by the very largest earthquakes and since
their impact on loss of life and properties is huge, it is of great importance to 
constrain as much as possible the shape of their distribution. We present a simple
and powerful probabilistic theoretical approach that shows that the Gamma
distribution is the best model, under the two hypothesis that the 
Gutenberg-Richter power law  distribution holds in absence of any condition and that
one or several constraints are imposed, either based on 
conservation laws or on the nature of the observations themselves. The selection of the Gamma
distribution does not
depend on the specific nature of the constraint. We illustrate the approach with two constraints,
the existence of a finite moment release rate and 
the observation of the size of a maximum earthquake in a finite catalog. Our
predicted ``soft'' maximum magnitudes compare favorably with those obtained by Kagan [1997] for the 
Flinn-Engdahl regionalization of subduction zones, collision zones and mid-ocean ridges.

\pagebreak

\section{Introduction}

The Gutenberg-Richter law states that the number $N(m)$ of earthquakes with
magnitude $\geq m$ is
\be
\log_{10} N(m) = a -b~m~,~~~~~~~~~{\rm with}~~b \approx 1~.
\ee
This relationship is best understood by transforming it
into a power law distribution for the
scalar seismic moment $M$ (expressed in $Nm$)
\be
P(M) dM = {\mu~M_t^{\mu} \over M^{1+\mu}}~dM~,~~~~~~
{\rm with}~~\mu \equiv {b \over \beta}~,~~~
{\rm for}~~M_t \leq M < +\infty~,
\label{rffgxb}
\ee
using the relation
\be
m = \frac{1}{\beta} [\log_{10} M - 9 ]~,
\label{jkgklg} 
 \ee
where $\beta$ is generally taken equal to $1.5$. $M_t$ is a lower 
seismic moment cut-off.

It is usually asserted that this self-similar law (\ref{rffgxb}) 
cannot be true for $M = + \infty$
because it would require that an infinite amount of energy be
released from the Earth's interior [{\it Wyss}, 1973; {\it Knopoff and Kagan},
1977; {\it Kagan}, 1994], since the mathematical
expectation or average $\langle M \rangle \equiv \int_{M_t}^{+\infty} dM~M~P(M) = +\infty$.

However, there is a subtlety not always appreciated in the literature that warrants
clarification. In fact, the statement that the self-similar law (\ref{rffgxb}) 
cannot be true for $M = + \infty$ due to the infinite expectation is not correct.
The point is that, notwithstanding an infinite mathematical expectation,
the measurement of the cumulative sum of seismic moments over any finite time interval $T$ will
always give a finite result! Thus, the infinite mathematical 
expectation does not require an infinite
amount of cumulative energy released from the Earth's interior and does
not contradict any conservation law. However, the infinite mathematical expectation 
is associated to the prediction that the cumulative sum of seismic moments should 
increase typically like the power $T^{1/\mu}= T^{3/2}$ of the time interval $T$ over
which it is measured. In other words, the rate of seismic moment release is predicted to
increase with time as $T^{{1 \over \mu}-1}$, with in addition huge fluctuations 
or jumps when a next biggest event occurs (which is precisely the origin of 
the acceleration of the rate). To sum up, the relevant question is not that of a finite 
or infinite expectation but rather whether the rate of cumulative growth of 
seismic moment is constant or increasing with observation time. Until this is 
completely settled observationally, the finiteness of the 
mathematical expectation of the seismic moments has the status of an hypothesis,
however plausible it may be.

A number of authors have remarked that the change from the two-dimensional
character of the fracture surface of small earthquakes to its one-dimensional character
for large earthquakes should lead to a modification of the exponent of the
Gutenberg-Richter law from a value $\mu \approx 2/3$ to a value larger than one (thus
ensuring convergence of the mean seismic rate), while
still keeping a power law shape [{\it Kanamori and Anderson}, 1975; {\it Rundle}, 1989; 
{\it Romanowicz}, 1992; {\it Pacheco et al.}, 1992; {\it Romanowicz and Rundle},
1993; {\it Okal and Romanowicz}, 1994; {\it Sornette and Sornette}, 1994; {\it Sornette et al.},
1996]. In these models, the cross-over between the two power laws 
is a measure of the thickness of the seismogenic zone in the case of strike-slip
earthquakes and of the downdip dimension of rupture in the case of
earthquakes in subduction zones. 

Another model assumes a sharp cut-off at a maximum moment $M_{max}$ such that absolutely
no earthquakes are to be expected with $M>M_{max}$
[{\it Anderson and Luco}, 1983], a version of which is used for instance in [{\it WGCEP}, 1995]
for the assessment of seismic hazards in southern California. 

Kagan [1993, 1994, 1997] and Main [1996]
have advocated a different ``soft'' cut-off according to which the power law is 
multiplied by an exponential roll-off at large moments\,:
\be
P_{xg}(M) ~dM= {C~M_t^{\mu} \over M^{1+\mu}}~\exp[-{M \over M_{xg}}]~dM~,
~~~~~{\rm for}~~M_t \leq M < +\infty~,
\label{rffgxqqqb}
\ee
where $C$ is a constant of normalization.
We will refer to this expression as a Gamma law.
Beyond $M_{xg}$, $P_{xg}(M)$ decays much faster than the pure Gutenberg-Richter law $P(M)$.
However, this is a ``soft'' cut-off since moments larger than $M_{xg}$ are possible but 
with smaller and smaller probabilities compared to the ``pure'' Gutenberg-Richter power law.
Quantitatively, the probability to exceed the corresponding magnitude $m_{xg}$ by a small fraction
of a unit magnitude is very small. For instance, the probability to exceed
$M_{xg}$ is $\int_{M_{xg}}^{\infty} P_{xg}(M) ~dM = 0.11 ~(M_{xg}/M_t)^{-\mu}$ (for $\mu=2/3$), i.e.
is smaller by a factor of about ten than the probability predicted from the extrapolation of the pure
power law distribution (\ref{rffgxb}). The probability to exceed
$7M_{xg}$, i.e. a magnitude $m=m_{xg}+0.6$ is $\int_{7M_{xg}}^{\infty} P_{xg}(M) ~dM = 
7~10^{-5} ~(7M_{xg}/M_t)^{-\mu}$ (for $\mu=2/3$), i.e.
is about ten thousand times less probable than the probability 
predicted from the extrapolation of the pure
power law distribution (\ref{rffgxb}). 

Since the tectonic seismic moment balance is dominated by the very largest earthquakes and since
their impact on loss of life and properties is huge, it is of great importance to 
constrain as much as possible the shape of their distribution. Here, we present a simple
but powerful and general probabilistic theoretical approach that shows that the Gamma
distribution is indeed the best model, under the following hypothesis\,: 
\begin{enumerate}
\item the unconditional 
Gutenberg-Richter power law  distribution (\ref{rffgxb}) is supposed to hold in the absence of
any constraint. This assumption cannot be verified in an absolute sense
 by any finite set of
observations which may present inherent constraints.
\item One or several constraints are imposed, either based on 
conservation laws (total observed tectonic deformation rate)
or on the nature of the observations themselves (maximum observed magnitude in a finite
time window). 
\end{enumerate}
Our approach uses a maximum likelihood formulation which can be interpreted using
information theory as a minimization of the information added from the 
specification of the constraints, given an earthquake catalog.
In a sense, it has similarities with the approach of Main and Burton [1984],
whose goal was to derive the Gutenberg-Richter distribution from 
the maximization of an information entropy. While the general Von Neumann/Shannon information
entropy principle is solidly based on fundamental probability theory, the choice of the 
specific entropy made in Main and Burton [1984] 
obviously controls the (power law) nature of the solution. This is spelled out in
detail by Shen and Mansinha [1983], who show that the probability distribution
generated from the principle of maximum entropy will depend on the (arbitrary) choice
of the independent variable and on the choice of the prior distribution representing
our complete ignorance. This prior distribution is shown to be inversely proportional
to the measurement errors, that in general are hard to estimate reliably.

In contrast, we do not claim
that the Gutenberg-Richter law derives from such a principle and we do not attempt to derive it
in any other way (see {\it Main}, 1996; {\it Grasso and Sornette}, 1998 for reviews of 
proposed physical mechanisms behind the Gutenberg-Richter law).  
Our goal is to quantify the distorsion imposed on
the Gutenberg-Richter law by the constraints.
 
 We first present the method in general terms and exemplify it on the simple case of
a dice play. We then apply it to the Gutenberg-Richter law and analyze two different
 types of constraints, namely the existence of a finite moment release rate and 
the observation of the size of the maximum moment in a finite catalog.

\section{Nature of the problem}

Formally, the question we address is how to find a probability distribution
satisfying the two conditions 1 and 2 of the previous section.

As pointed by V. Pisarenko (private communication), a natural way to address
this question is to use some functional 
$R(P ; P_x)$ measuring the ``distance'' between the Gutenberg-Richter density $P$ and the
desired modified density $P_x$. Then, one can minimize this distance $R(P ; P_x)$
under the given conditions. The problem is that there is a certain degree
of arbitrariness in the choice of the distance $R(P ; P_x)$, that lead to different
solutions.

If one takes the Kullback Distance 1 
representing average log-likelihood  $\ln [P_x (v)/P(v)]$  of $P_x$  
against $P$ [{\it Kullback}, 1958]
\be
R_1(P ; P_x) = \int dv~ P_x(v) ~ \ln [P_x(v)/P(v)]   ~,
\label{eq111}
\ee
one gets directly (using the Lagrange multiplier method) the Gamma distribution
\be
P_x(v) = P(v) ~ e^{a - bv}~,
\label{kjqkql}
\ee
where the constants $a$ and $b$ are determined by the constraints. 

However, there are other ``distances'' that are a priori as justifiable as the 
Kullback Distance 1 and which lead to different results.
The Kullback Distance 2 is the average log-likelihood of $P$ against $P_x$
\be
R_2(P ; P_x) = \int dv~ P(v) ~ \ln [P(v)/P_x(v)]    ~,
\label{dgfgg}
\ee
which leads to the modified solution
\be
P_x(v) = {P(v) \over a + bv}~,
\ee
where the constants $a$ and $b$ are again determined by the constraints. 

Another example is the Kullback Distance 3, quantifying 
the ``divergence'' between  $P$ and $P_x$
\be
R_3(P ; P_x ) = R_1(P ; P_x) + R_2(P ; P_x)~,
\ee
which leads to the following solution
\be
P_x(v) = {P(v) \over G(a + bv)}~,
\ee
where $G(z)$ is the inverse function of $g(z) = z + ln z$.

Thus, as a consequence of the existence of some
degree of arbitrariness in the choice of distance $R(P ; P_x)$, this approach
does not select the Gamma law (\ref{kjqkql}) as the unique solution of
the two conditions 1 and 2 of the previous section.

Our approach provides a way to avoid the arbitrariness in the choice of distance
between $P$ and $P_x$, based on fixing the random sample mean of observed
events. This constraint could be seen as too restrictive, since it selects among
all realizations of possible sequences of earthquakes only those where the
sample mean is exactly equal to a specified value. Our point is that this 
constraint is fixed to a specific value by an independent global measurement
of the cumulative strain obtained by geodetic or satellite techniques, thus providing
an estimation of the cumulative released moment (neglecting for the time
being difficulties associated with the tensor nature of the problem).
Thus, we propose to condition the modified Gutenberg-Richter
distribution on those specific random
realizations that are consistent with the global measurement. 
This rather specific constraint would not apply in all circumstances.

We now turn to a presentation of this approach.

\section{Frequencies conditioned by constraints}

Let $v_1, v_2, ..., v_{n-1}, v_n$ be the different possible values of a random
variable distributed according to a given a priori density distribution $P(v)$.
In the earthquake case, if magnitudes
are given with a precision $0.1$, they can only take values such as 
$...,5.6, 5.7, 5.8, 5.9, 6.0, 6.1...$. In general,
experimental or observational data are always coarse-grained at a given resolution level, thus
making the measurements discrete. In the sequel, we use discrete notations, as it is 
straightforward to write our results in a continuous framework by simply taking 
the limit of infinite resolution.

A given catalog consists of $N$ events. Each event, say the $l$-th one, has a value $v_l$ taken 
from the pool of the $n$ possible values. Then, the
classical law of large numbers [{\it Gnedenko and Kolmogorov}, 1954]
 states that the frequency $f(v_l)$ with
which one measures a given value  $v_l$ among $n$ possible values 
of the same random variable converges towards its a priori probability $P(v_l)$ as $N \to \infty$.
This idea is in fact at the basis of the frequency concept of probability. 

Now, suppose that the measured mean of
these $N$ realizations deviates from the theoretical mean. What can we say about the
frequency of each value $v_l$? More generally, suppose that we possess additional observational
constraints. What are the consequences for the relative frequencies of the events?

This question has a well-defined mathematical answer in the limit of large $N$. 
In general, the frequency of a given value $v_l$
converges to a well-defined number in the limit of large $N$  but this number $f(v_l)$ is
different from the probability $P(v_l)$! The fundamental reason for this result is that there is
a close relationship between the existence of the deviation of the mean from its
theoretical value and the existence of frequencies that are different from their
theoretical probabilities. Probability theory allows one to compute
precisely this anomalous behavior in the limit of large $N$. 
Our results also hold for finite $N$.
Technically, this analysis belongs to the so-called
large deviation regime [{\it Lanford}, 1973; {\it Frisch}, 1995].

Let $y_1, ..., y_N$ be the $N$ observed values that can take one of the $n$ possible
values $v_1, ..., v_n$. We denote the observed mean $V$ by
\be
 V \equiv {1 \over N} \sum_{j=1}^N y_j~,
\ee
 which can also be written  
\be
 V =  \sum_{l=1}^n f_l v_l ,
 \label{zae}
\ee
where $f_l = {N_l \over N}$ is the observed frequency of the value $v_l$, 
$N_l$ is the total number of realizations and $N$ is the total
number of events. In expression (\ref{zae}), the $N$
realizations have been partitioned into groups of identical values, so that the first
group contains $N_1$ variables each equal to $v_1$, the second group contains $N_2$
variables each equal to $v_2$, and so on. Therefore, $N_1 + N_2 + ... + N_n = N$.

The law of large numbers 
 states that $f_l = {N_l \over N}  \rightarrow P(v_l)$, 
when $N \rightarrow \infty$. For an
observed value of $V = x$, what can we say about $f_l$? More
precisely, what are the values taken by $f_l$, conditioned by the observation of $V = x$?
In the earthquakes application, the constraint on the average may
for instance reflect the long-term tectonic deformation balance. 
A by-product of our calculation will be an assessment of the impact 
on the earthquake frequency distribution of an error made in the
estimation of the long-term tectonic deformation or of its possible non-stationarity.

To solve this problem, we first estimate the probability to observe the frequencies
$f_1, f_2, ..., f_n$ from a total of $N$ realizations of the random variable\,: 
\be
P(f_1, f_2, ..., f_n) = {N! \over (Nf_1)! (Nf_2)! ...(Nf_n)!}  \prod_{l=1}^n
[P(v_l)]^{Nf_l} ~ ,
\label{feawvbx}
\ee
where $Nf_l = N_l$ and $(Nf_l)!$ is $Nf_l (Nf_l-1) (Nf_l-2) ...
4.3.2.1$. To write (\ref{feawvbx}), we have used the assumption that the events
are independent. Using the Stirling formula $\ln N! \approx N \ln N - N$
to expand $(Nf_l)!$, we find
\be
P(f_1, f_2, ..., f_n) \simeq e^{- N H(f_1, f_2, ..., f_n)} ~,
\label{wxcqsea}
\ee
where
\be 
H(f_1, f_2, ..., f_n) = \sum_{l=1}^n f_l ~\log{f_l \over P(v_l)}
\label{aezrdfd}
\ee
is often called neguentropy (the negative of the entropy).
For $N$ large, the law of large numbers states that
the frequencies $f_l$ converge towards the values that minimize
$H(f_1, f_2, ..., f_n)$ in the presence of constraints (i.e. such that 
$P(f_1, f_2, ..., f_n)$ is maximum). Note that, while $f_1, f_2, ..., f_n$
are random values, they converge to well-defined non-random values 
$f(v_1), f(v_2), ..., f(v_n)$ for $N \to + \infty$.

In the absence of constraints, the normalization condition
$\sum_{l=1}^n f_l = 1$ reduces (\ref{wxcqsea}) to the law of large number
\be
f_l \rightarrow  P(v_l) ~.
\ee

However, if we observe $V \equiv
\sum_{l=1}^n f_l v_l = x$, the maximum likelihood 
frequencies are those that minimize the function
\be
H(f_1, f_2, ..., f_n) - \lambda_1 \biggl[\sum_{l=1}^n f_l -1\biggl]  -\lambda_2
\biggl[\sum_{l=1}^n f_l v_l - x \biggl]~,
\ee
where $\lambda_1$ and $\lambda_2$ are two
Lagrange multipliers. The technique of Lagrange multipliers allows one to
solve an optimization problem in the presence of constraints
by incorporating them in the function to be minimized.
Then, the solution depends on these factors, which
are then eliminated by imposing the constraints on the solution [{\it Bertsekas}, 1982].
In the present case, the two
constraints associated to $\lambda_1$ and $\lambda_2$ are the 
normalization of the $f_j$'s and the observation that $V=x$.
The solution is
\be 
f_l \rightarrow f(v_l) = P(v_l) ~{e^{-\beta v_l} \over Z(\beta)} ~ , 
\label{rytold}
\ee
where $\beta(x) \equiv - \lambda_2$, which is determined as a function of $x$ from
\be
{d \log Z(\beta) \over d \beta} = -x~,
\label{conciicic}
\ee
where the ``partition function'' $Z(\beta)$ is defined by 
\be
Z(\beta) = \sum_{l=1}^n P(v_l) ~e^{-\beta v_l} .
\label{zzzzzdh}
\ee
The expression (\ref{rytold}) gives the frequencies of the values of the random
variables $v_l$, conditioned by the existence of a constraint on the mean. 
Notice that the case $\beta = 0$ retrieves the unconstrained case. We thus expect
that the behavior of $Z(\beta)$ close to $\beta = 0$ controls most situations
where the constraints introduce only minor perturbations in most of the
distribution.

It is not fortuitous that the
expressions (\ref{rytold}-\ref{zzzzzdh}) bear a strong
similarity with the statistical mechanics formulation [{\it Rau}, 1997] of
systems composed of many elements, where 
$Z(\beta)$ is the partition function, $\beta$ is the inverse
temperature and $-\log Z(\beta)$ is proportional to the free energy.
The fact that the constraint is seen as a ``high temperature'' ($\beta \to 0$)
perturbation is clear\,: the constraint is analogous to an ``energy'' added
to the ``free energy'' $-\log Z(\beta)$, which in the absence of constraint is solely controlled
by the ``entropy'' $H$. The relative importance of entropy and energy is weighted by 
the temperature, the entropy dominating at high temperatures.

Let us illustrate this result (\ref{rytold}) on a dice game. Suppose that a dice with six faces
is thrown $N$ times and that one counts the frequencies 
$f_l$, $l=1$ to $n=6$ with which each of the six
faces of the dice occur. According to the law of large numbers, the six $f_l$ tend
to $1/6 \simeq 0.166$ for large $N$ and the mean $f_1 + 2 f_2 + 3 f_3
+ 4 f_4 + 5 f_5 + 6 f_6$ tends to $3.5$. 
Let us now assume that we observe a mean $x = 4$. The formula 
(\ref{rytold}) predicts that the frequencies that contribute to this deviation
are not the same anymore. We get $f_1 \approx 0.103$, $f_2 \approx
0.123$,  $f_3 \approx 0.146$,  $f_4 \approx 0.174$, $f_5 \approx
0.207$ and $f_6 \approx 0.247$. In other words, the five and six occur about
twice as much as the one. Intuitively, the large
values become more frequent because the outcomes are biased by
the observation of a larger mean. 

We stress that this probabilistic approach does not explain {\it why} the constraint exists or
why there is a deviation from the mean. It 
simply draws the best conclusions on the frequencies that are compatible with the existence or
observation of such a constraint.

\section{Application to earthquakes}

\subsection{Constraint of a finite total moment release\,: theory of the Gamma
distribution}

Consider the normalized power law distribution $P(v)$
\be
P(v) ~dv  = {\mu \over v^{1+\mu}}~dv~,
\label{guyr}
\ee
where $v = M/M_t$ is the normalized seismic moment and $\mu \approx 2/3$.
From this normalization, the integral from 
$1$ to $\infty$ of $P(v)$ is equal to $1$.

It is convenient to return to continuous notation, for which expression 
(\ref{zzzzzdh}) for $Z(\beta)$ reads
\be
Z(\beta) = \int_1^{\infty} dv~{\mu ~e^{-\beta v} \over v^{1+\mu}} = 
\mu \beta^{\mu} \int_{\beta}^{\infty} dx {e^{-x} \over x^{1 + \mu}} ~.
\label{olknjkl}
\ee
In the appendix, we show that for large $x$, i.e. small $\beta$
\be
Z(\beta) = e^{-d_{\mu} \beta^{\mu}}~, 
\label{laplaformds}
\ee
with 
\be
d_{\mu} = \Gamma(1-\mu)~,
\label{ttggg}
\ee
where $\Gamma$ denotes the Gamma function.
Condition (\ref{conciicic}) leads to 
\be
\beta_a (x) = \biggl({\mu \Gamma(1-\mu) \over x}\biggl)^{1 \over 1-\mu}~.
\label{bbbdhdhj}
\ee
The observed frequency of earthquake size is thus
\be
f_l \rightarrow f(v_l) = C(\mu, x){ e^{-\beta_a(x) v_l} \over v_l^{1+\mu}} ~,
\label{resulsx}
\ee
where
\be
C(\mu,x) = \mu ~\exp \biggl( \Gamma(1-\mu) [\beta_a(x)]^\mu \biggl)~.
\ee
The expression (\ref{resulsx}) recovers the Gamma distribution (\ref{rffgxqqqb})
postulated by Kagan [1993, 1994, 1997]. Using $v = M/M_t$, we rewrite (\ref{resulsx}) as 
(\ref{rffgxqqqb}) where the soft ``cut-off'' $M_{xg}$ is given by
\be
M_{xg} \equiv {M_t \over \beta_a(x)} = M_t ~
\biggl({x \over {\mu \Gamma(1-\mu)}}\biggl)^{1 \over 1-\mu}~.
\label{jkggg}
\ee

Let us translate this formula into geologically meaningful quantities. Here,
we follow the notation of Kagan [1997] and will test our results against his.
The quantity $x$ can be expressed
as a function of the geological rate $\dot{M}$ of deformation (in $Nm/year$)
of the region under consideration and of the yearly number $\alpha_t = N/\Delta t$ 
of events with moments above the threshold $M_t$ (taken from the Harvard
catalog over its lifespan $\Delta t = 18.5$~years) as follows
\be
x = {\dot{M}~\Delta t \over M_t~N} ~.
\label{kglmdgkds}
\ee
Using $\beta = 2/3$, which we shall assume fixed, we get from (\ref{jkggg})
\be
M_{xg} = M_t ~ ({x \over 1.786})^3~,
\label{kgllg}
\ee
and
\be
m_{xg} = m_t + 2 ~\log_{10} {x \over 1.786}~,
\label{gfhdj}
\ee
where $m_{xg}$ is the magnitude corresponding to $M_{xg}$ and $m_t$ is the magnitude
corresponding to the moment threshold $M_t$. This equation is essentially the same as the one used by
Kagan [1997]. In particular, (\ref{gfhdj}) with (\ref{kglmdgkds}) shows that 
$m_{xg} = 2 ~\log_{10} \dot{M}  + $~constant, in agreement with Kagan [1997, his equation (9)] for
$\mu = 2/3$. The only difference between our equation and Kagan's equation
(9) is in the additive factors and in our fixed choice for 
$\mu = 2/3$. Fixing $\mu$ is justified by the fact that some catalogs 
are very short and it is reasonable to limit as much as possible the number of free
variables. From (\ref{gfhdj}), we see that an error on $x$, i.e. on $\dot{M}$,
of a factor two results into an error of $0.6$ in the ``soft cut-off'' magnitude $m_{xg}$.

Following Kagan [1997], we take 
$m_t = 5.8$ corresponding to $M_t = 10^{17.7}~Nm = 0.5~10^{18}~Nm$.
The results are summarized in the Tables 1-3 for subduction zones, collision zones and 
mid-ocean ridges. We compare our estimation for the ``soft cut-off'' magnitude $m_{xg}$ 
for the Gamma law derived from our analysis (\ref{gfhdj}) with the corresponding 
values obtained by Kagan [1997] for the Flinn-Engdahl seismic regions. We find very good 
agreement with the estimations of Kagan [1997] for the subduction zones and collision zones.
Our estimation of $m_{xg}$ for mid-ocean ridges remains in the range $8-9$ also in agreement 
with Kagan [1997] who quotes the value $m_{xg} = 8.7$, assuming $\mu=0.63$. Our results 
for mid-ocean ridges are however
less sensitive to large fluctuations to unphysical values [{\it Kagan}, 1997].

The expressions (\ref{jkggg},\ref{gfhdj}) with (\ref{kglmdgkds}) show that $M_{xg}$ and 
$m_{xg}$ increase with $x$, i.e. with the constraining geological rate of deformation $\dot{M}$.
If $\dot{M}$ goes to infinity (the constraint no longer exists), 
we recover the pure Gutenberg-Richter distribution with
$M_{xg} \to \infty$. The relation $M_{xg} \sim \dot{M}^{1 \over 1-\mu}$ obtained from
(\ref{jkggg}) follows from this simple argument. $\dot{M} \sim N \langle M \rangle$ is the sum of 
$N$ earthquake contributions,
where $\langle M \rangle$ is the average seismic moment released per earthquake. 
For a power law distribution with an effective truncation at $M_{xg}$, 
$\langle M \rangle \sim \int_{M_t}^{M_{xg}} dM~M/M^{1+\mu} 
\sim M_{xg}^{1-\mu}$. Inverting $\dot{M} \sim M_{xg}^{1-\mu}$, we get the dependence of
$M_{xg}$ as a function of $\dot{M}$ in (\ref{jkggg}).

Is the statistical estimate
of the curvature of the Gutenberg-Richter law based on ten or twenty earthquakes 
reliable? In our experience, we have found that formulas based on 
asymptotic large limits often work satisfactory even for remarkably small systems 
that would a priori rule out their validity. Let us mention, for instance, continuous
hydrodynamics equations that work in superfluid helium flows for fluid layer thicknesses equal 
to a fraction of the size of one helium atom (see for instance
Noiray et al. [1984]). Another example is the liquid-solid transition of
clusters of atoms (see for instance Matsuoka et al. [1992])
that has essentially all its infinite limit thermodynamic properties 
as soon as the number of atoms is larger than a few tens. There are many other examples where
extrapolation of asymptotic formulas valid for large statistics provide (good) surprises in the 
small statistical limit.

\subsection{Constraint of a finite total moment release and of a maximum
size}

Let us now assume that the average seismic moment release is again $V = x$ 
(in normalized units) and in addition 
no observations have shown $v > v_{max}$. In other words, $M_{max} = M_t~v_{max}$ is the largest 
seismic moment observed in the catalog.
How are the observable frequencies $f(v)$
deviating from the ``pure'' pdf $P(v)$ (still assumed to be the pure power law 
(\ref{guyr}))? One could argue that this constraint is not natural since it might
simply result from the artifact of a limited catalog. But this is precisely the
question we ask\,: conditioned by the absence of $v$'s larger than $v_{max}$, what is
the implication for the distorsion of the derived distribution? Here, we are simply
pointing out that as long as a ``great'' earthquake does not occur, this may lead to the
false estimation that the distribution is truncated, while in fact the truncation, if any,
occurs at larger unsampled values.

The answer is obtained by following the steps in the previous section. This leads
to the modified expression of the partition function
\be
Z(\beta) = \int_1^{v_{max}} dv~{\mu ~e^{-\beta v} \over v^{1+\mu}}~.
\ee
This can be written as
$$
Z(\beta) = \int_1^{\infty} dv~{\mu ~e^{-\beta v} \over v^{1+\mu}}
-  \int_{v_{max}}^{\infty} dv~{\mu ~e^{-\beta v} \over v^{1+\mu}}
$$
\be
= \int_1^{\infty} dv~{\mu ~e^{-\beta ~v} \over v^{1+\mu}}
- [v_{max}]^{-\mu} \int_{1}^{\infty} dv~{\mu ~e^{-\beta ~v_{max}~v} \over v^{1+\mu}}~.
\ee
Proceeding as in the previous section and using the result in the appendix,
we get from (\ref{laplaformds})
\be
Z(\beta) = e^{-d_{\mu} \beta^{\mu}} -  [v_{max}]^{-\mu}~e^{-d_{\mu} [\beta v_{max}]^{\mu}}~.
\label{laplaqdffq}
\ee
The inverse ``temperature'' $\beta$ is again given by (\ref{conciicic}).

Two cases must be considered.
\begin{enumerate}
\item $\beta_a(x)~ v_{max} > 1$\,: this condition is the same as $M_{max} > M_{xg}$, i.e. the largest
observed seismic moment is larger that the ``soft cut-off'' $M_{xg}$ of the Gamma distribution.
In this case, 
the second term in (\ref{laplaqdffq}) can be neglected and we recover the previous results (\ref{bbbdhdhj}).
The impact of $M_{max}$ is negligible.
\item $\beta_a(x)~ v_{max} < 1$\,: the observation of the maximum observed magnitude will modify
the observed Gutenberg-Richter law as we now calculate.
\end{enumerate}

In the case $\beta_a(x)~ v_{max} < 1$, the simplest approach is to expand (\ref{laplaqdffq}) in
powers of $\beta^{\mu}$ up to second order. Then,
 as the first-order cancels out between the first and second
term in the r.h.s.\,:
\be 
Z(\beta) \approx (1- [v_{max}]^{-\mu}) - {d_{\mu}^2 \over 2} ~v_{max}^{\mu}~\beta^{2\mu}~,
\ee
which yields
\be
-{d \log Z(\beta) \over d \beta} \approx \mu~d_{\mu}^2 ~v_{max}^{\mu}~\beta^{2\mu-1}~.
\ee
Equating to $x$ according to the equation (\ref{conciicic}) yields
\be
\beta(x) = \biggl( {x \over \mu ~[\Gamma(1-\mu)]^2~v_{max}^{\mu}}\biggl)^{1 \over 2\mu-1} = 
\biggl({x \over 4.784 ~v_{max}^{\mu}} \biggl)^3~,
\ee
using $\mu = 2/3$. Inverting, we get the modified ``soft cut-off'' seismic moment entering
into the Gamma distribution 
\be
M_{xg}^{max} = M_t ~ \biggl({4.784 ~v_{max}^{2 \over 3} \over x}\biggl)^3~.
\label{kgllqqqqqg}
\ee
This has a dependence in $x$ which is the inverse of the previous case (\ref{kgllg})
where the impact of the observation of the largest earthquake is not felt.

As a case study, assume that the ``soft cut-off'' magnitude $m_{xg}$ is about $8.5$.
Using (\ref{gfhdj}), this corresponds to a value $x=40$. This magnitude is
the most probable value found in the results shown in the Tables 1-3 as well as by Kagan [1997].
Let us assume that we have only a limited catalog and that the largest earthquake in this catalog has
a magnitude $m_{max} = 7.5$, which corresponds to $v_{max} = M_{max}/M_t = 355$.
 This example corresponds to Southern California with the largest 
earthquake in the southern earthquake catalog being the Kern county 1952 earthquake.
Introducing these values  $x=40$ and $v_{max} = 355$ in (\ref{kgllqqqqqg}) yields
$M_{xg}^{max} = 1.08~10^{20}~Nm$, i.e. $m_{xg}^{max} = 7.36$. This theory thus predicts
a slight bending down of the Gutenberg-Richter law at a value slightly smaller than
the maximum observed magnitude. This occurs when this later value is significantly less than the 
``soft cut-off'' magnitude $m_{xg}$ solely deduced by the geological tectonic deformation rate.
In this example, the number of earthquakes of magnitude close to $7.4$ is about one third
the number that would be extrapolated from the pure Gutenberg-Richter power law.
If the maximum observed magnitude is $m_{max} = 7.0$ corresponding to $v_{max} = 316$, we get
$M_{xg}^{max} = 0.86~10^{20}~Nm$, i.e. $m_{xg}^{max} = 7.28$.

\subsection{Other constraints}

Another example is motivated by the recent debate
as to whether there is a deficit in intermediate size earthquake in southern California since 1850
[{\it WGCEP}, 1995; {\it Stein and Hanks}, 1998]. One can estimate the
modified Gutenberg-Richter law induced by both this deficit and the constraint of
a finite moment release by using the same technique as described above. Let us briefly
indicate how to proceed. In the simplest version, we consider a total deficit between $v_1$ and $v_2$,
i.e. no events of size $v_1 \leq v \leq v_2$ occurred. 
We still keep the constraint of a given average moment release.
The solution is obtained by following the same steps as in the previous section, which lead
to the modified expression of the partition function
\be
Z(\beta) = \int_1^{v_1} dv~{\mu ~e^{-\beta v} \over v^{1+\mu}} +
\int_{v_2}^{\infty} dv~{\mu ~e^{-\beta v} \over v^{1+\mu}}~,
\ee
which can also be written as
$$
Z(\beta) = \int_1^{\infty} dv~{\mu ~e^{-\beta v} \over v^{1+\mu}}
-  \int_{v_1}^{\infty} dv~{\mu ~e^{-\beta v} \over v^{1+\mu}}
+ \int_{v_2}^{\infty} dv~{\mu ~e^{-\beta v} \over v^{1+\mu}}~.
$$
\be
= \int_1^{\infty} dv~{\mu ~e^{-\beta ~v} \over v^{1+\mu}}
- [v_1]^{-\mu} \int_{1}^{\infty} dv~{\mu ~e^{-\beta ~v_1~v} \over v^{1+\mu}}
+ [v_2]^{-\mu} \int_{1}^{\infty} dv~{\mu ~e^{-\beta ~v_2~v} \over v^{1+\mu}}~.
\ee
Using the result of the appendix, we get from (\ref{laplaformds})
\be
Z(\beta) = e^{-d_{\mu} \beta^{\mu}} -  [v_1]^{-\mu}~e^{-d_{\mu} [v_1\beta]^{\mu}}
+ [v_2]^{-\mu}~e^{-d_{\mu} [v_2\beta]^{\mu}}~.
\label{laplaqdggddffq}
\ee
The inverse ``temperature'' $\beta$ is again given by (\ref{conciicic}).  Different cases
can then be analyzed as a function of the relative influence of the values $v_1$, $v_2$ and
the global geological deformation rate $x$.

\section{A physical derivation of the Gamma distribution}

In this section, we complement the analysis by proposing a simple physically-based derivation of 
the Gamma distribution. Our previous considerations have been of a probabilistic nature. It is
useful to extend our intuition by identifying the structure of physical models that are
compatible with a Gamma distribution. The simple model discussed now shows that the self-similarity
and homogeneity conditions, together with a self-consistent cascade mechanism, lead 
to the Gamma distribution.

In the spirit of reaction rate theory and in analogy with various approaches to model
fragmentation processes [{\it Cheng and Redner}, 1988; {\it Brown and Wohletz}, 1995], 
we posit that the number of earthquakes of ``energy'' $v$ is
the solution of the following self-consistent integral equation\,:
\be
P(v) = C \int_v^{+\infty} dv'~P(v')~f(v' \to v)~,
\label{gjklkjh}
\ee
where $C$ is a constant and $f(v' \to v)$ is the distribution of event sizes $v$ arising from
a cascade of earthquakes triggered by the event of size $v'$. This model envisions that 
the crust self-organizes into a state where events are correlated, each of them being able to trigger
a set of smaller earthquakes (aftershocks). The lower bound $v$ in the integral express that 
earthquakes are preferentially triggered by larger preceeding events.
We assume that $f(v' \to v)$ is a power law of $v'/v$ with exponent $1+\mu$ 
($f(v' \to v) = (v'/v)^{1+\mu}$). Then, (1) the power law assumption will 
lead to the Gutenberg-Richter distribution\,; (2) the homogeneous dependence in $v'/v$ is the simplest
law compatible with self-similarity. The solution of (\ref{gjklkjh}) is found to be the Gamma distribution\,:
\be
P(v) = P_0~{1 \over v^{1+\mu}}~e^{-{v \over v_{max}}} ~,
\ee
where $P_0$ is a normalizing constant.

The general class of branching models [{\it Vere-Jones}, 1977]
provides a simple concrete geometrical implementation of this cascade model (\ref{gjklkjh}). 
Branching models describe the notion of
a cascade that may either end after a finite number of steps or diverge, depending
upon the value of a control parameter, the branching probability. 
It has been applied to describe failure and earthquakes, seen as resulting from
a succession of events chained through a causal connection [{\it Vere-Jones}, 1977].
The resulting distribution of event size is the Gamma distribution in the sub-critical regime.

\section{Concluding remarks}

We have shown how to formulate the effect of a global constraint on the observed
Gutenberg-Richter law, using simple probability concepts. The remarkable result is that
a constraint leads in general to a modification of the Gutenberg-Richter power law
into a Gamma law, as advocated by Kagan  [1993, 1994, 1997], with a
``soft cut-off'' magnitude controlled by the constraint. This provides a strong basis for the
use of the Gamma distribution as a model of earthquake frequency-moment distribution.

Technically, the reason for this result may be seen to lie in the fact that
the logarithmic density of frequencies normalized by $N$ given by
expression (\ref{aezrdfd}) is exactly like the Kullback Distance 1 given by
expression (\ref{eq111}). In this sense, our approach proposes in this context an original 
way to solve for the a priori arbitrariness in the choice of the
distance between distributions. One could thus state that the Gamma-distribution
gets still another justification.

This general approach can be used to study the impact on the Gutenberg-Richter law
stemming from the existence of other observational constraints. We have thus also shown how to 
incorporate the observational constraint of the existence of a maximum observed magnitude
and have outlined how the observation of a deficit of earthquakes in a certain magnitude
window could also be tackled by this technique.

\vskip 0.5cm
We acknowledge useful discussions with D.D. Jackson, Y.Y. Kagan and G. Ouillon and especially 
illuminating correspondence with V.F. Pisarenko. We also thank I. Main as 
a referee for thoughtful remarks and J. Pujol as the associate editor for a careful 
reading of the manuscript.

\pagebreak
\noindent APPENDIX

\noindent We wish to calculate $Z(\beta)$ given by 
(\ref{olknjkl}). Let us be general and consider the case where $\mu$ can be larger than $1$. This
situation has been argued for large earthquakes [{\it Pacheco et al.}, 1992; {\it 
Sornette et al.}, 1996].
Denote $l$ the integer part of $\mu$ ($l<\mu<l+1$). Integrating by part $l$
times, we get 
$$
Z(\beta) = e^{-\beta} \biggl(1 - {\beta \over \mu - 1} + ... + {(-1)^l
\beta^l \over (\mu - 1)(\mu - 2)...(\mu - l)} \biggl) + 
$$
\be
+{(-1)^l \beta^{\mu} \over (\mu - 1)(\mu - 2)...(\mu - l)}
 \int_{\beta}^{\infty} dx e^{-x} x^{l - \mu} ~.
\ee
This last integral is equal to
\be 
\beta^{\mu} \int_{\beta}^{\infty} dx e^{-x} x^{l - \mu} =
\Gamma(l+1-\mu) [\beta^{\mu} + \beta^{l+1} \gamma^*(l+1-\mu,\beta)]~ ,
\ee
where $\Gamma$ is the Gamma function ($\Gamma(n+1)=n!$ for an integer argument)
and 
\be
\gamma^*(l+1-\mu,\beta)=
e^{-\beta} \sum_{n=0}^{+\infty} {\beta^n \over \Gamma(l+2-\mu+n)}
\ee
is the incomplete Gamma function [{\it Abramowitz and Stegun}, 1972]. 
We see that $Z(\beta)$ 
presents a regular Taylor expansion in powers of $\beta$ up to the order
$l$, followed by a term of the form $\beta^\mu$.
We can thus write
\be
Z(\beta) = 1 + r_1 \beta + ..... + r_l \beta^l +
 r_\mu \beta^{\mu} + {\cal O}(\beta^{l+1}) ~,
\ee
with $r_1 = -\langle v \rangle, \ r_2={\langle v^2 
\rangle \over 2}, ...$.. For small $\beta$, 
we rewrite $Z(\beta)$ under the form
\be
Z(\beta) = \exp\left[-\sum_{k=1}^l d_k \beta^k - d_\mu \beta^{\mu} \right]~,  
\label{laplaform}
\ee
where the coefficient $d_k$ can be simply expressed in terms of the $r_k$'s.
In particular, we have
\be
d_{\mu} = \Gamma(l+1-\mu)~.
\label{dddsss}
\ee
The expression (\ref{laplaform}) generalizes the canonical form
 of the characteristic function of the stable L\'evy laws [{\it Samorodnitsky and Taqqu}, 1994]
for arbitrary values of $\mu$, and not solely for $\mu \leq 2$ for which they are
defined. L\'evy laws can exist only for $\mu<2$ and are stable  upon convolution, 
i.e. their shape is unchanged
up to a rescaling. The generalized expression (\ref{laplaform}) shows
that the {\it tail} of a power law distribution also remains stable even for $\mu > 2$. However,
the tail slowly shrinks in size as the domain of validity of the power law tail for $\mu > 2$
extends beyond a limit which slowly increases as $\sqrt{N \ln N}$ with the number $N$ of events.

Note that the canonical form of the characteristic function of L\'evy laws
is recovered for $\mu \leq 2$ for which the coefficient
$d_2$ is not defined (the variance does not exist) and the only analytical
term is $\langle v \rangle \beta$ (for $\mu > 1$).

For the earthquake application, $\mu < 1$ and we thus obtain
\be
Z(\beta) = e^{-d_{\mu} \beta^{\mu}}~~~~~~~~{\rm for~small}~~\beta,~~{\rm i.e.~large}~~x~. 
\ee

\pagebreak
{\bf References}

Abramowitz, M. and I.A. Stegun (1972). {\ it Handbook of 
mathematical functions}, Dover publications, New York.

Bertsekas, D.P. (1982). {\it Constrained optimization and Lagrange multiplier methods}, New York : 
Academic Press.

Brown, W.K., and K.H. Wohletz (1995). Derivation of the Weibull distribution based on physical 
principles and its connection to the Rosin-Rammler and lognormal distributions, 
{\it J. Appl. Phys.}, {\bf 78}, 2758-2763.

Cheng, Z., and S. Redner (1988). Scaling theory of fragmentation,
{\it Phys. Rev. Lett.}, {\bf 60}, 2450-2453.
    
Frisch, U. (1995). {\it Turbulence, The legacy of A.N. Kolmogorov},
Cambridge University Press, Cambridge.

Gnedenko, B.V., and A.N. Kolmogorov (1954). {\it Limit distributions
for sum of independent random variables}, Addison Wesley, Reading MA.

Grasso, J.R., and D. Sornette (1998). Testing self-organized criticality by induced seismicity, 
{\it J. Geophys.Res.}, {\bf  103}, 29965-29987.
 
Kagan, Y. Y. (1993). Statistics of characteristic earthquakes, 
{\it Bull. Seismol. Soc. Am.}, {\bf 83}, 7-24.

Kagan, Y. Y. (1994). Observational evidence for earthquakes as a nonlinear dynamic
process, {\it Physica D}, {\bf 77}, 160-192.

Kagan, Y.Y. (1997). Seismic moment-frequency relation for shallow earthquakes: Regional
comparison, {\it J. Geophys. Res.}, {\bf 102}, 2835-2852.

Kanamori, H., and D. L. Anderson (1975).
Theoretical basis of some empirical relations in seismology,
{\it Bull. Seismol. Soc. Am.}, {\bf 65}, 1073-1095.

Knopoff, L., and Y. Y. Kagan (1977). Analysis of the theory of extremes as
applied to earthquake problems, {\it J. Geophys. Res.}, {\bf 82}, 5647-5657.

Kullback, S. (1958). {\it Information Theory and Statistics},
John Wiley, New York, Chapter 1.

Lanford, O.E. (1973). {\it Entropy and equilibrium states in classical
mechanics}, in Statistical Mechanics and Mathematical Problems,
Lect.Notes in Physics 20, 1-113, ed. A. Lenard, Springer, Berlin.

Main, I. (1996). Statistical physics, seismogenesis and seismic hazard, {\it Reviews of
Geophysics}, {\bf 34}, 433-462.
 
Main, I.G., and P.W. Burton (1984). Information theory and the earthquake 
frequency-magnitude distribution, {\it Bull. Seis. Soc. Am.}, {\bf 74}, 1409-1426.
  
Matsuoka, H., T. Hirokawa, M. Matsui and M. Doyama (1992). Solid-liquid transitions in Argon
clusters, {\it Phys. Rev. Lett.}, {\bf 69}, 297-300

Noiray, J.C., D. Sornette, J.P.  Romagnan and J.P. Laheurte (1984).
Stratification Transition In He3-He4 Mixture Films, {\it Phys. Rev. Lett.}, {\bf 53}, 2421-2424.

Okal, E. A., and B. A. Romanowicz (1994). On the variation of
$b$-values with earthquake size,{\it Phys. Earth Planet. Inter.},{\bf 87}, 55-76.

Pacheco, J. F., C. H. Scholz, and L. R. Sykes (1992).
Changes in frequency-size relationship from small to large
earthquakes, {\it Nature}, {\bf 355}, 71-73.

Rau, J. (1997). Statistical Mechanics in a Nutshell, preprint physics/9805024,
Lecture Notes (Part I of a course on ``Transport Theor'' taught at 
Dresden University of Technology, Spring 1997.

Romanowicz, B. (1992). Strike-slip earthquakes on quasi-vertical transcurrent
faults: Inferences for general scaling behavior,
{\it Geophys. Res. Lett.}, {\bf 19}, 481-484.

Romanowicz, B., and J. B. Rundle (1993). On scaling relations for
large earthquakes, {\it Bull. Seismol. Soc. Am.}, {\bf 83}, 1294-1297.

Rundle, J. B. (1989). Derivation of the complete Gutenberg-Richter
magnitude-frequency relation using the principle of scale
invariance, {\it J. Geophys. Res.}, {\bf 94}, 12,337-12,342.

Samorodnitsky, G., and M. S. Taqqu (1994). {\it Stable non-Gaussian random processes:
stochastic models with infinite variance}, New York : Chapman \& Hall.  

Shen, P.Y., and L. Mansinha (1983). On the principle of maximum entropy and the
earthquake frequency-magnitude relation, {\it Geophys. J. R. Astr. Soc.}, {\bf 74},
777-785.

Sornette, D., and A. Sornette (1994). On scaling relations for large earthquakes 
from the perspective of a recent nonlinear diffusion equation linking short-time 
deformation to long-time tectonics-Comment, {\it Bull.Seism.Soc.Am.}, {\bf 84}, 1679-1683. 

Sornette, D., L. Knopoff, Y.Y. Kagan and C. Vanneste (1996).
Rank-ordering statistics of extreme events : application to the distribution of
large earthquakes, {\it J.Geophys.Res.}, {\bf 101}, 13883-13893.

Stein, R.S., and T.C. Hanks (1998). $M \geq 6$ earthquakes in southern California during the 
twentieth century: no evidence for a seismicity or moment deficit, 
{\it Bull. Seismol. Soc. Am.}, {\bf 88}, 635-652.

Vere-Jones, D. (1977). Statistical theories of crack propagation, {\it Mathematical
Geology}, {\bf 9}, 455-481.

Working group on California earthquake probabilities, Seismic hazards in southern
California: probable earthquakes, 1994 to 2024 (1995). {\it Bull. Seism. Soc. Am.}, {\bf 85}, 379-439.

Wyss, M. (1973). Towards a physical understanding of the earthquake frequency distribution, 
{\it Geophys. J. R. Astr. Soc.}, {\bf 31}, 341-359.

\pagebreak

\begin{center}
\centering{Table 1\\
SUBDUCTION ZONES\,: Soft ``cut-off'' magnitude $m_{xg}$ for the Gamma law derived from 
equation (\ref{gfhdj}) and comparison to Kagan [1997] fits (indicated by 
$[m_{xg} \pm \sigma_m]^{a}$ for Flinn-Engdahl seismic regions. 
$N$ is the number of earthquakes in each region with seismic moment larger than the threshold $M_t$.}
\end{center}
 
\begin{table*}[h]
\begin{center}
\begin{tabular}{|c|c|c|c|c|c|} \hline
Seismic Region &  $N$ & $\dot{M}$ & $x$ & $[m_{xg} \pm \sigma_m]^{a}$  & $m_{xg}$ (Eq.(\ref{gfhdj})) \\ \hline
Alaska-Aleutian Arc & 152 &  $1.80~10^{20}~Nm$ & $44$ & $8.53 \pm 0.29$ & $8.58$\\ \hline
Mexico-Guatemala & 83 &  $0.84~10^{20}~Nm$ & $37$ & $8.41 \pm 0.30$ & $8.44$\\ \hline
Central America & 86 &  $0.88~10^{20}~Nm$ & $38$ & $8.42 \pm 0.30$ & $8.45$\\ \hline
Caribbean Loop & 31 &  $0.37~10^{20}~Nm$ & $44$ & $8.54 \pm 0.34$ & $8.58$\\ \hline
Andean S. America & 125 &  $3.00~10^{20}~Nm$ & $89$ & $9.09 \pm 0.29$ & $9.19$\\ \hline
Kermadec-Tonga-Samoa & 248 &  $2.10~10^{20}~Nm$ & $31$ & $8.27 \pm 0.28$ & $8.29$\\ \hline
Fiji Is. & 44 &  $0.81~10^{20}~Nm$ & $68$ & $8.88 \pm 0.31$ & $8.96$\\ \hline
New Hebrides Is. & 222 &  $1.70~10^{20}~Nm$ & $28$ & $8.19 \pm 0.28$ & $8.20$\\ \hline
Bismarck-Solomon Is. & 219 &  $1.74~10^{20}~Nm$ & $29$ & $8.22 \pm 0.28$ & $8.23$\\ \hline
New Guinea & 129 &  $3.00~10^{20}~Nm$ & $85$ & $9.07 \pm 0.29$ & $9.16$\\ \hline
Guam-Japan & 43 &  $1.02~10^{20}~Nm$ & $88$ & $9.08 \pm 0.32$ & $9.18$\\ \hline
Japan-Kamchatka & 227 &  $3.00~10^{20}~Nm$ & $49$ & $8.62 \pm 0.28$ & $8.67$\\ \hline
S.E. Japan-Ryukyu Is. & 22 &  $0.64~10^{20}~Nm$ & $117$ & $9.24 \pm 0.36$ & $9.35$\\ \hline
Taiwan & 52 &  $0.54~10^{20}~Nm$ & $38$ & $8.43 \pm 0.31$ & $8.46$\\ \hline
Philippines & 147 &  $1.25~10^{20}~Nm$ & $31$ & $8.27 \pm 0.29$ & $8.29$\\ \hline
Borneo-Celebes & 149 &  $1.47~10^{20}~Nm$ & $36$ & $8.39 \pm 0.29$ & $8.42$\\ \hline
Sunda Arc & 122 &  $2.30~10^{20}~Nm$ & $70$ & $8.90 \pm 0.29$ & $8.98$\\ \hline
Adaman Is.-Sumatra & 26 &  $0.94~10^{20}~Nm$ & $133$ & $9.41 \pm 0.35$ & $9.55$\\ \hline

\end{tabular}
\end{center}
\end{table*}

 \pagebreak
 
 \begin{center}
\centering{Table 2\\
COLLISION ZONES\,: Soft ``cut-off'' magnitude $m_{xg}$ for the Gamma law derived from 
equation (\ref{gfhdj}) and comparison to Kagan [1997] fits (indicated by 
$[m_{xg} \pm \sigma_m]^{a}$ for Flinn-Engdahl seismic regions. 
$N$ is the number of earthquakes in each region with seismic moment larger than the threshold $M_t$.}
\end{center}
 
\begin{table*}[h]
\begin{center}
\begin{tabular}{|c|c|c|c|c|c|} \hline
Seismic Region &  $N$ & $\dot{M}$ & $x$ & $[m_{xg} \pm \sigma_m]^{a}$  & $m_{xg}$ (Eq.(\ref{gfhdj})) \\ \hline
Burma-S.E. Asia & 20 &  $0.52~10^{20}~Nm$ & $96$ & $9.15 \pm 0.37$ & $9.26$\\ \hline
India-Tibet-Yunan & 29 &  $0.45~10^{20}~Nm$ & $57$ & $8.75 \pm 0.34$ & $8.81$\\ \hline
S. Sinkiang-Kansu & 20 &  $0.21~10^{20}~Nm$ & $39$ & $8.44 \pm 0.37$ & $8.47$\\ \hline
W. Asia & 50 &  $0.38~10^{20}~Nm$ & $28$ & $8.18 \pm 0.32$ & $8.19$\\ \hline
M.-E.-Crimea-Balkans & 37 &  $0.29~10^{20}~Nm$ & $29$ & $8.21 \pm 0.33$ & $8.22$\\ \hline
W. Mediterranean & 22 &  $0.15~10^{20}~Nm$ & $25$ & $8.10 \pm 0.36$ & $8.10$\\ \hline
Baluchistan & 10 &  $0.37~10^{20}~Nm$ & $137$ & $9.43 \pm 0.45$ & $9.57$\\ \hline
\end{tabular}
\end{center}
\end{table*}

\vskip 2cm
 
 \begin{center}
\centering{Table 3\\
MID-OCEANS RIDGES\,: Soft ``cut-off'' magnitude $m_{xg}$ for the Gamma law derived from 
equation (\ref{gfhdj}) and comparison to Kagan [1997] fits (indicated by 
$[m_{xg} \pm \sigma_m]^{a}$ for Flinn-Engdahl seismic regions. 
$N$ is the number of earthquakes in each region with seismic moment larger than the threshold $M_t$.}
\end{center}
 
\begin{table*}[h]
\begin{center}
\begin{tabular}{|c|c|c|c|c|c|} \hline
Seismic Region &  $N$ & $\dot{M}$ & $x$ & $[m_{xg} \pm \sigma_m]^{a}$  & $m_{xg}$ (Eq.(\ref{gfhdj})) \\ \hline
Baja-Calif-Gulf Calif. & 10 &  $0.17~10^{20}~Nm$ & $63$ & $12.81 \pm 2.45$ & $8.89$\\ \hline
Atlantic Ocean & 112 &  $0.67~10^{20}~Nm$ & $22$ & $8.28 \pm 1.61$ & $7.98$\\ \hline
Indian Ocean & 94 &  $1.44~10^{20}~Nm$ & $57$ & $12.36 \pm 1.63$ & $8.80$\\ \hline
Artic & 8 &  $0.15~10^{20}~Nm$ & $69$ & $13.24 \pm 2.64$ & $8.98$\\ \hline
S.E. \& Antartic Pacific & 107 &  $2.38~10^{20}~Nm$ & $82$ & $13.98 \pm 1.61$ & $9.12$\\ \hline
Galapagos & 16 &  $1.44~10^{20}~Nm$ & $332$ & $20.05 \pm 2.15$ & $10.34$\\ \hline
Macquarie Loop & 48 &  $0.43~10^{20}~Nm$ & $33$ & $10.03 \pm 1.74$ & $8.33$\\ \hline

\end{tabular}
\end{center}
\end{table*}

\end{document}